\documentclass[final]{ias2}

\usepackage{graphicx}

\begin{document}

\title{The Hoyle state in Nuclear Lattice EFT}

\author[aa]{Timo A. L\"ahde}
\email{t.laehde@fz-juelich.de}
\author[ab]{Evgeny Epelbaum}
\author[ab]{Hermann Krebs}
\author[ac]{Dean Lee}
\author[aa,ad,ae]{Ulf-G.~Mei{\ss}ner}
\author[af]{Gautam Rupak}
\address[aa]{Institute~for~Advanced~Simulation, Institut~f\"{u}r~Kernphysik, and 
J\"{u}lich~Center~for~Hadron~Physics, Forschungszentrum~J\"{u}lich, D--52425~J\"{u}lich, Germany}
\address[ab]{Institut~f\"{u}r~Theoretische~Physik~II,~Ruhr-Universit\"{a}t~Bochum, D--44870~Bochum,~Germany}
\address[ac]{Department~of~Physics, North~Carolina~State~University, Raleigh, NC~27695, USA}
\address[ad]{Helmholtz-Institut f\"ur Strahlen- und Kernphysik and Bethe Center for Theoretical Physics,
Universit\"at Bonn, D--53115 Bonn, Germany}
\address[ae]{JARA~--~High~Performance~Computing, Forschungszentrum~J\"{u}lich, D--52425 J\"{u}lich,~Germany}
\address[af]{Department~of~Physics~and~Astronomy, Mississippi~State~University, Mississippi State, MS~39762, USA}

\begin{abstract}
We review the calculation of the Hoyle state of $^{12}$C in Nuclear Lattice Effective Field Theory (NLEFT) and its anthropic implications for the
nucleosynthesis of $^{12}$C and $^{16}$O in red giant stars. We also review the extension of NLEFT to the regime of medium-mass nuclei, with
emphasis on the determination of the ground-state energies of the alpha nuclei $^{16}$O, $^{20}$Ne, $^{24}$Mg and $^{28}$Si by means of Euclidean
time projection. Finally, we review recent NLEFT results for the spectrum, electromagnetic properties, and alpha-cluster structure of $^{16}$O.
\end{abstract} 


\maketitle


\section{Introduction}

Nuclear Lattice Effective Field Theory (NLEFT) is a first-principles approach, in which Chiral EFT for nucleons is combined with 
numerical Auxiliary-Field Quantum Monte Carlo (AFQMC) lattice calculations. NLEFT differs from other \textit{ab initio} 
methods~\cite{Hagen:2012fb,Jurgenson:2013yya,Roth:2011ar,Hergert:2012nb,Lovato:2013cua,Duguet} in that it is an 
unconstrained Monte Carlo calculation, which does not rely on truncated basis expansions or many-body perturbation theory, nor on prior 
information about the structure of the nuclear wave function.

As in Chiral EFT, our calculations are organized in powers of a generic soft scale $Q$ associated with factors of momenta and the 
pion mass~\cite{Epelbaum:2008ga}. 
We denote $\mathcal{O}(Q^0)$ as leading order (LO), $\mathcal{O}(Q^2)$ as next-to-leading order (NLO), and $\mathcal{O}(Q^3)$
as next-to-next-to-leading order (NNLO) contributions. The present calculations are performed up to NNLO.
We define $H_\mathrm{LO}^{}$ as the LO lattice Hamiltonian, and $H_\mathrm{SU(4)}^{}$ as the equivalent Hamiltonian with
the pion-nucleon coupling $g_A^{} = 0$ and contact interactions that respect Wigner's SU(4) symmetry. 

In our NLEFT calculations, $H_\mathrm{LO}^{}$ is treated non-perturbatively (see Ref.~\cite{Dean_QMC} for a review).
The NLO contribution to the two-nucleon force (2NF), the electromagnetic and strong isospin-breaking contributions (EMIB), 
and the three-nucleon force (3NF) which first enters at NNLO, are all treated as perturbations. It should be noted that our ``LO'' calculations 
use smeared short-range interactions that capture much of the corrections usually treated at NLO and higher orders~\cite{Borasoy:2006qn}.
At NNLO, the 3NF overbinds 
nuclei with $A \geq 4$ due to a clustering instability which involves four nucleons on the same lattice site. A long-term objective of NLEFT is to
remedy this problem by decreasing the lattice spacing and including the next-to-next-to-next-to-leading order (N3LO) 
corrections in Chiral EFT. In the mean time, the overbinding problem
has been rectified by means of a 4N contact interaction, tuned to the empirical binding energy of either $^4$He or 
$^8$Be~\cite{Epelbaum:2009pd}. While this provides a good description of the alpha nuclei up to $A = 12$ including the Hoyle 
state~\cite{Epelbaum:2009pd,Epelbaum:2011md,Epelbaum:2012qn}, the overbinding is found to increase more rapidly for $A \geq 16$. 
In Ref.~\cite{A28_letter}, a non-local 4N interaction which accounts for all possible configurations of four nucleons on adjacent lattice sites was
introduced, and adjusted to the empirical binding energy of $^{24}$Mg.
A detailed study of the spectrum and electromagnetic properties of $^{16}$O (with inclusion of the effective 4N interaction) 
has been reported in Ref.~\cite{16O_spectrum}.


\section{The Hoyle state}

The Hoyle state is a resonance with spin-parity quantum numbers $J^p = 0^+$ in the spectrum of $^{12}$C, which plays an important role in resonantly enhancing
the reaction rate for the so-called triple-alpha process, which is responsible for the production of carbon in massive stars that have reached the red giant
stage in their evolution. This reaction represents a significant bottleneck in the stellar nucleosynthesis, as $^8$Be is an unstable (though relatively long-lived)
resonance. For $^{12}$C to form, a third alpha particle must combine with the $^8$Be resonance to form the Hoyle state, which subsequently decays
electromagnetically to the ground state of $^{12}$C. This reaction may then proceed further (non-resonantly) to form $^{16}$O through addition of a fourth
alpha particle. However, the temperature of the stellar plasma at which the triple-alpha process takes place depends exponentially on the energy 
$\Delta_h^{}$ of the
Hoyle state above the triple-alpha threshold, which is experimentally known to be $\Delta_h^{} \simeq 379.5$~keV. Stellar model 
calculations~\cite{Oberhummer} have shown 
that only a narrow window of $\pm 100$~keV exists in $\Delta_h^{}$ where sizable amounts of carbon and oxygen can be produced simultaneously.

\begin{table}[b]
\centering
\begin{tabular}{c|c|c} 
& \hspace{.4cm} LO \hspace{.4cm} 
& \hspace{.4cm} Exp \hspace{.4cm} \\ \hline\hline
$r(0_{1}^{+})$ [fm] & $2.2(2)$ & $2.47(2)$ \cite{Schaller:1982} \\
$r(2_{1}^{+})$ [fm] & $2.2(2)$ & $-$ \\
$Q(2_{1}^{+})$ [$e\,$fm$^{2}$] & $6(2)$ & $6(3)$ \cite{Vermeer:1983} \\
$r(0_{2}^{+})$ [fm] & $2.4(2)$ & $-$ \\
$r(2_{2}^{+})$ [fm] & $2.4(2)$ & $-$ \\
$Q(2_{2}^{+})$ [$e\,$fm$^{2}$] & $-7(2)$ & $-$
\end{tabular}
\caption{Lattice results at leading order (LO) and available experimental values for the
root-mean-square charge radii and quadrupole moments of the $^{12}$C
states. \label{Hoyle1}}
\end{table}

Our NLEFT calculations~\cite{Epelbaum:2012qn} 
have recently shed light on the structure, electromagnetic properties and transitions of the Hoyle state from first principles, and have
furthermore allowed for a first investigation of the sensitivity of the triple-alpha process to changes in the fundamental parameters, such as the light quark
mass $m_q^{}$ and the electromagnetic fine-structure constant $\alpha_\mathrm{em}^{}$. In particular, carbon-oxygen based life as we
know appears unlikely to become strongly disfavored for relative changes smaller than $\simeq 3$\% in $m_q^{}$ or $\alpha_\mathrm{em}^{}$. This situation,
where the production rates of $^{12}$C and $^{16}$O are not excessively fine-tuned, 
arises due to strong correlations between the binding energy of $^4$He and the energies of the $^8$Be and Hoyle state
resonances. In turn, this is a reflection of the underlying alpha-cluster structure of the Hoyle state, which we have found to resemble a ``bent-arm'' or
obtuse triangular configuration. However, such calculations also require as input some knowledge on how the LO contact terms in the Chiral EFT 
interaction depend on $m_q^{}$ or, equivalently, the pion mass 
$m_\pi^{}$. At present, this information enters through the derivatives of the two-nucleon $S$-wave scattering lengths in the singlet and triplet channels, 
$\partial a_s^{-1}/\partial m_\pi^{}$ and $\partial a_t^{-1}/\partial m_\pi^{}$ respectively. These have
so far proven difficult to determine accurately from Lattice QCD
calculations. In this situation, the derivatives of the binding energies $\partial B_3^{}/\partial m_\pi^{}$ and $\partial B_4^{}/\partial m_\pi^{}$ of 
$^3$He and $^4$He may prove more constraining in the near term, as these are easier to extrapolate to the physical point from
Lattice QCD data~\cite{WIMP}. NLEFT work in this direction is in progress.

\begin{table}[h]
\centering
\begin{tabular}{l|c|c} 
& \hspace{.4cm} LO \hspace{.4cm} 
& \hspace{.4cm} Exp \hspace{.4cm} \\ \hline\hline
$B(E2,2_{1}^{+}\rightarrow0_{1}^{+})$ [$e^{2} $fm$^{4}$] & $5(2)$ & $7.6(4)$
\cite{Ajzenberg:1990a} \\
$B(E2,2_{1}^{+}\rightarrow0_{2}^{+})$ [$e^{2} $fm$^{4}$] & $1.5(7)$ &
$2.6(4)$ \cite{Ajzenberg:1990a} \\ 
$B(E2,2_{2}^{+}\rightarrow0_{1}^{+})$ [$e^{2} $fm$^{4}$] & $2(1)$ &
$-$ \\
$B(E2,2_{2}^{+}\rightarrow0_{2}^{+})$ [$e^{2} $fm$^{4}$] & $6(2)$ &
$-$ \\ 
$M(E0,0_{2}^{+}\rightarrow0_{1}^{+})$ [$e\,$fm$^{2}$] & $3(1)$ & $5.5(1)$
\cite{Chernykh:2010zu}
\end{tabular}
\caption{Lattice results at leading order (LO) and available experimental values for
electromagnetic transitions involving the even-parity states of $^{12}$C.
\label{Hoyle2}}
\end{table}

Finally, we briefly review our results for the electromagnetic properties of the low-lying even-parity states of $^{12}$C in Table~\ref{Hoyle1},
and our results for the electromagnetic transitions between these states in Table~\ref{Hoyle2}. These are currently available
to LO (the extension to NNLO will appear in a future publication). We note that the good agreement with experimental results (where available)
inspires confidence in our conclusions concerning the alpha-cluster structure and possible anthropic role of the Hoyle state.


\section{Medium-mass nuclei in NLEFT}

Recently, much effort has been directed towards the extension of NLEFT beyond $^{12}$C into the regime of medium-mass 
nuclei (see in particular Ref.~\cite{A28_letter}). In this section, we shall review the developments in computational
methods which have made this extension possible. At present, our calculations have been performed with a (spatial) lattice spacing 
of $a=1.97$~fm in a periodic cube of length $L = 11.8$~fm. In Ref.~\cite{A28_letter}, our trial wave function
$|\Psi_{A}^\mathrm{init}\rangle$ is a Slater-determinant state composed of delocalized standing waves,
with $A$ nucleons and the desired spin and isospin. First, we 
project $|\Psi_{A}^\mathrm{init}\rangle$ for a time $t^\prime$ using the Euclidean-time evolution operator of the 
SU(4) Hamiltonian, giving the ``trial state'' 
$|\Psi_A^{}(t^\prime_{})\rangle \equiv \exp(-H_\mathrm{SU(4)}^{} t^\prime_{}) |\Psi_{A}^\mathrm{init}\rangle$.
Second, we use the full Hamiltonian $H_\mathrm{LO}^{}$ to construct the Euclidean-time projection amplitude
\begin{equation}
Z_A^{}(t) \equiv \langle\Psi_A^{}(t^\prime_{})| \exp(-H_\mathrm{LO}^{} t) |\Psi_A^{}(t^\prime_{})\rangle,
\label{ZAt}
\end{equation}
and the ``transient energy'' 
\begin{equation}
E_A^{}(t) = -\partial[\ln Z_A^{}(t)]/\partial t.
\label{EAt}
\end{equation}
If we denote
by $|\Psi_{A,0}^{}\rangle$ the lowest (normalizable) eigenstate of $H_\mathrm{LO}^{}$
which has a non-vanishing overlap with the trial state $|\Psi_A^{}(t^\prime_{})\rangle$,
we obtain the corresponding energy $E_{A,0}^{}$ as the ${t\to\infty}$ limit of $E_A^{}(t)$.
The NLO and NNLO contributions are evaluated in perturbation theory. We compute
operator expectation values using
\begin{equation}
Z_A^\mathcal{O}(t) \equiv \langle\Psi_A^{}(t^\prime_{})| \exp(-H_\mathrm{LO}^{} t/2)
\mathcal{O} \exp(-H_\mathrm{LO}^{} t/2)  |\Psi_A^{}(t^\prime_{})\rangle,
\label{OP}
\end{equation}
for any operator $\mathcal{O}$. Given the ratio $X_A^\mathcal{O}(t) = Z_A^\mathcal{O}(t)/Z_A^{}(t)$, the expectation value 
of $\mathcal{O}$ for the desired state $|\Psi_{A,0}^{}\rangle$ is obtained as 
$X_{A,0}^\mathcal{O} \equiv \langle\Psi_{A,0}^{}| \mathcal{O} |\Psi_{A,0}^{}\rangle = \lim_{t \to \infty}X_A^\mathcal{O}(t)$.

\begin{figure}[t]
\includegraphics[width=\columnwidth]{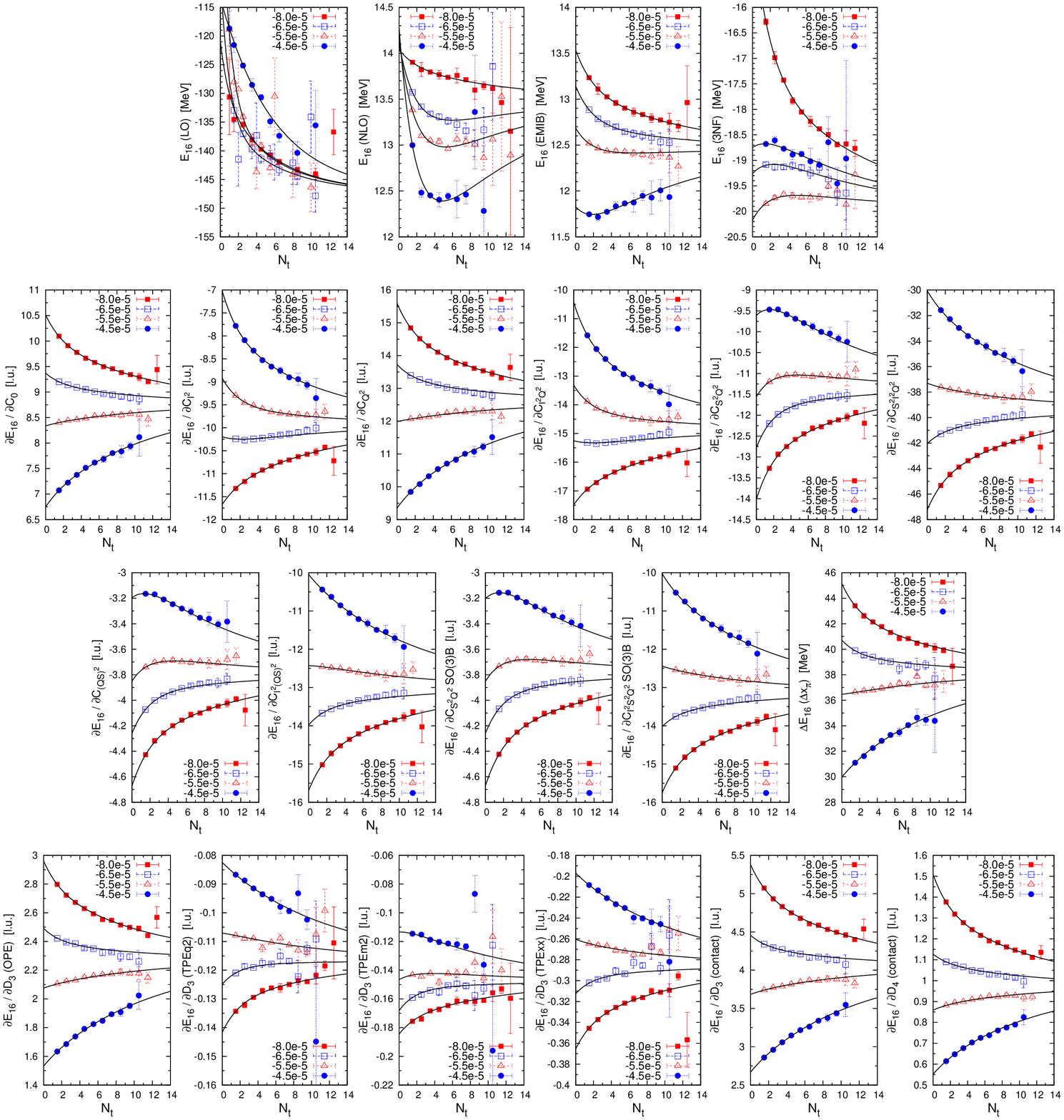}
\vspace{-.7cm}
\caption{NLEFT results for $^{16}$O. The LO energy is $E_\mathrm{LO}^{} = -147.3(5)$~MeV, and the result
at NNLO including 4N interactions is $E_\mathrm{NNLO+4N}^{} = -131.3(5)$~MeV. The empirical binding energy is
$-127.62$~MeV. For definitions, see the main text.
\label{16O}}
\end{figure}

Sign oscillations make it difficult to reach sufficiently large values of the 
projection time~$t$. It is helpful to note that the closer the trial state $|\Psi_A^{}(t^\prime_{})\rangle$ is to $|\Psi_{A,0}^{}\rangle$, 
the less the necessary projection time $t$. 
$|\Psi_A^{}(t^\prime_{})\rangle$ can be optimized by adjusting both the SU(4) projection time $t^\prime_{}$ and the strength 
of the coupling $C_\mathrm{SU(4)}^{}$ of $H_\mathrm{SU(4)}$. 
The accuracy of the extrapolation $t\to\infty$ can be further improved by simultaneously incorporating data from trial states that differ in 
$C_\mathrm{SU(4)}^{}$. 
The large-time behavior of $Z_A^{}(t)$ and $Z_A^{\mathcal{O}}(t)$ is controlled by the low-energy spectrum of $H_\mathrm{LO}^{}$. 
Let $| E\rangle$ label the eigenstates of $H_\mathrm{LO}^{}$ with energy $E$, and let $\rho_{A}^{}(E)$ denote the density of states for a system
of $A$~nucleons. We then express $Z_A^{}(t)$ and $Z_A^{\mathcal{O}}(t)$ in terms of their spectral representations,
\begin{align}
Z_A^{}(t) = & \int dE \: \rho_A^{}(E) \:
\big| \langle E |\Psi_A^{}(t^\prime_{})\rangle\big|^2_{} 
\exp(-Et), \\
Z_A^{\mathcal{O}}(t) = & \int dE\,dE^\prime_{} \, \rho_A^{}(E)\,\rho_A^{}(E^\prime_{}) 
\langle\Psi_A^{}(t^\prime_{})|E\rangle \,
\langle E|\mathcal{O}|E^\prime_{}\rangle 
\nonumber \\
& \langle E^\prime_{}|\Psi_A^{}(t^\prime_{})\rangle \, \exp(-(E+E^\prime_{})t/2),
\end{align} 
from which we construct the spectral representations of $E_A^{}(t)$ and $X_A^{\mathcal{O}}(t)$. 
We can approximate these to arbitrary accuracy over any finite range of $t$ by taking $\rho_{A}^{}(E)$ to be a sum of energy delta functions,
$\rho_{A}^{}(E) \approx \sum_{i=0}^{i_\mathrm{max}}\delta(E-E_{A,i}^{})$,
where we take $i_\mathrm{max} = 4$ for the $^4$He ground state, 
and $i_\mathrm{max} = 3$ for $A \geq 8$. Using data obtained for
different values of $C_\mathrm{SU(4)}^{}$, we perform a correlated fit of $E_A^{}(t)$ and $X_A^{\mathcal{O}}(t)$ for all operators $\mathcal{O}$
that contribute to the NLO and NNLO energy corrections. We find that the use of $2-6$ trial states allows for a much more precise
determination of $E_{A,0}^{}$ and $X_{A,0}^{\mathcal{O}}$ than hitherto possible. In particular, we may ``triangulate'' 
$X_{A,0}^{\mathcal{O}}$ using trial states that correspond to functions $X_A^{\mathcal{O}}(t)$ which converge both from above and below.


\section{Lattice Monte Carlo results for NLEFT with $A \geq 16$}

In this section, we discuss our NLEFT results for nuclei in the medium-mass range (see also Refs.~\cite{A28_letter,16O_spectrum}). 
A detailed description of the NLEFT calculation
for $^{16}$O is given in Fig.~\ref{16O}, while the results for the alpha nuclei from $^4$He to $^{28}$Si are shown in Table~\ref{tab_en}.
The curves in Fig.~\ref{16O} show a correlated fit for all trial states, using the same 
spectral density $\rho_A^{}(E)$. The upper row shows (from left to right) the LO energy, the total isospin-symmetric 
2NF correction (NLO), the electromagnetic and isospin-breaking corrections (EMIB) and the total 3NF correction. 
The remaining panels show the matrix elements $X_A^{\mathcal{O}}(t)$ that form part of the NLO and 3NF terms. 
The operators $\partial E_{A}^{}/\partial C_i^{}$ give the contributions of the NLO contact interactions, and 
$\Delta E_{A}^{} (\Delta x_\pi^{})$ denotes the energy shift due the $\mathcal{O}(a^2)$-improved pion-nucleon coupling. 
The operators $\partial E_{A}^{}/\partial D_i^{}$ give the individual contributions to the total 3NF correction. 

\begin{table}[h]
\begin{center}
\begin{tabular}{r | r | r r r | r}
$A$ & \multicolumn{1}{c |}{LO (2N)} & \multicolumn{1}{c}{NNLO (2N)} 
& \multicolumn{1}{c}{+3N} & \multicolumn{1}{c |}{+4N$_\mathrm{eff}$} &
\multicolumn{1}{c}{Exp} \\ \hline\hline
$4$ & $-28.87(6)$ & $-25.60(6)$ & $-28.93(7)$ & $-28.93(7)$ & $-28.30$ \\
$8$ & $-57.9(1)$ & $-48.6(1)$ & $-56.4(2)$ & $-56.3(2)$ & $-56.35$ \\
$12$ & $-96.9(2)$ & $-78.7(2)$ & $-91.7(2)$ & $-90.3(2)$ & $-92.16$ \\
$16$ & $-147.3(5)$ & $-121.4(5)$ & $-138.8(5)$ & $-131.3(5)$ & $-127.62$ \\
$20$ & $-199.7(9)$ & $-163.6(9)$ & $-184.3(9)$ & $-165.9(9)$ & $-160.64$ \\
$24$ & $-253(2)$ & $-208(2)$ & $-232(2)$ & $-198(2)$ & $-198.26$ \\
$28$ & $-330(3)$ & $-275(3)$ & $-308(3)$ & $-233(3)$ & $-236.54$\\
\end{tabular}
\caption{NLEFT results for the ground-state energies (in MeV).
The combined statistical and extrapolation errors are given in parentheses. The columns labeled ``LO(2N)'' and ``NNLO(2N)'' show
the energies at each order using the two-nucleon force only. The column labeled ``+3N'' also includes the 3NF, which first appears 
at NNLO. Finally, the column ``+4N$_\mathrm{eff}$'' includes the ``effective'' 4N force. The column ``Exp'' gives
the empirical energies.
\label{tab_en}}
\end{center}
\end{table}

From the results in Table~\ref{tab_en}, we note that the NNLO results are good up to $A = 12$, at which point an increasing
over-binding manifests itself for $A \geq 16$. 
As we ascend the alpha ladder from $^{4}$He to $^{28}$Si, the lighter nuclei can be described as collections of 
alpha clusters~\cite{Epelbaum:2012qn,Epelbaum:2011md}. As the number of clusters increases, they become increasingly densely packed, such that
a more uniform liquid of nucleons is approached. This increase in the density of alpha clusters appears correlated with the gradual overbinding 
we observe at NNLO for $A \geq 16$. As this effect becomes noticeable for $^{16}$O, we can view it as a problem which first
arises in a system of four alpha clusters. 

Following Ref.~\cite{Epelbaum:2009pd}, which removed discretization errors associated with four 
nucleons occupying the same lattice site, we can attempt to remove similar errors associated with four alpha clusters in close proximity on 
neighboring lattice sites. In Table~\ref{tab_en}, the column
labeled ``+4N$_\mathrm{eff}$'' shows the results at NNLO while including both the 3NF and the ``effective'' nearest-neighbor 4N interaction
$V^{(4\mathrm{N_{eff}})}$. Due to the low momentum cutoff (corresponding to a lattice spacing of $a = 1.97$~fm), the
two-pion exchange contributions have been absorbed into the contact interactions at NLO.
We have tuned the coupling $D^{(4\mathrm{N_{eff}})}$ of $V^{(4\mathrm{N_{eff}})}$
to give approximately the correct energy for the ground state of $^{24}$Mg. With 
$V^{(4\mathrm{N_{eff}})}$ included, a good description of the ground-state energies is obtained over the full range from light to medium-mass nuclei, with
a maximum error no larger than $\sim 3$\%. This lends support to the qualitative picture that the overbinding at NNLO in 
Table~\ref{tab_en} is associated with the increased packing of alpha clusters and the eventual crossover to a uniform nucleon liquid.  
The missing physics would then be comprised of short-range repulsive forces that counteract the dense packing of alpha clusters. 

\begin{table}[t]
\centering
\begin{tabular}{c | r | r r r | r}
$J_n^p$ & \multicolumn{1}{c |}{LO (2N)} & \multicolumn{1}{c}{NNLO (2N)} 
& \multicolumn{1}{c}{+3N} & \multicolumn{1}{c |}{+4N$_\mathrm{eff}$} & \multicolumn{1}{c}{Exp} \\ \hline\hline
$0^+_1$ & $-147.3(5)$ & $-121.4(5)$ & $-138.8(5)$ & $-131.3(5)$ & $-127.62$ \\
$0^+_2$ & $-145(2)$ & $-116(2)$ & $-136(2)$ & $-123(2)$ & $-121.57$ \\
$2^+_1$ & $-145(2)$ & $-116(2)$ & $-136(2)$ & $-123(2)$ & $-120.70$
\end{tabular}
\caption{NLEFT results and experimental (Exp) energies for the lowest even-parity states of $^{16}$O (in MeV). 
The errors are one-standard-deviation estimates which include statistical Monte Carlo errors and 
uncertainties due to the extrapolation $N_t^{} \to \infty$. 
\label{oxygen1}}
\end{table}

In spite of the good agreement (upon introduction of $V^{(4\mathrm{N_{eff}})}$) with experiment in Table~\ref{tab_en},
we also need to verify that this good description of the binding energies is not accidental.
It is then helpful to check whether a consistent
picture is obtained with respect to excited states, transitions and electromagnetic properties of nuclei in the medium-mass range where 
$V^{(4\mathrm{N_{eff}})}$ gives a sizable contribution. 


\section{Alpha-cluster structure of $^{16}$O}

Since the early work of Wheeler~\cite{Wheeler:1937zza}, theoretical studies of $^{16}$O have been based on alpha-cluster 
models~\cite{Dennison:1954zz,Iachello,Robson:1979zz,Bauhoff:1984zza,Filikhin:1999,Tohsaki:2001an,Bijker}, and some 
experimental evidence for alpha-particle substructure in $^{16}$O has been found from the analysis of decay products~\cite{Freer:2005ia}.  
While some of the puzzles in the structure of $^{16}$O have been explained on a phenomenological (or geometrical) level, so far 
no support has been available for the alpha-cluster structure of $^{16}$O from {\it ab initio} calculations.
Our NLEFT results for $^{16}$O have been reported in Ref.~\cite{16O_spectrum}, which is also
the first time that evidence for the tetrahedral alpha-cluster structure of the ground state of $^{16}$O has been found from an {\it ab initio} calculation. 
We have also found the first excited $0^+$ state of $^{16}$O to predominantly consist of a square arrangement of alpha clusters.
We summarize our results for the electromagnetic properties and transition rates in $^{16}$O in Tables~\ref{oxygen1} and~\ref{oxygen2}. 

The computed charge radii, quadrupole moments and transition rates of $^{16}$O provide very convincing evidence supporting the realism of our extension
of NLEFT to medium-mass nuclei. In particular, the excitation energies and level ordering in $^{16}$O (see Table~\ref{oxygen1}) is found to be very sensitive 
to the strength and form of $V^{(4\mathrm{N_{eff}})}$. This sensitivity arises due to the differences in the alpha-cluster structure of the states in question.
We also note that NLEFT is able to explain the empirical value of $B(E2,2^+_1 \to 0^+_2)$, which is $\simeq 30$ times larger than the Weisskopf 
single-particle shell model estimate (see Table~\ref{oxygen2}). This provides confirmation of the interpretation of the $2^+_1$ state as a rotational excitation of the 
$0^+_2$ state. Finally, we provide a prediction for the quadrupole moment of the $2^+_1$ state. We note that the NLEFT calculation of the 
electromagnetic transitions requires a full coupled-channel analysis.  For such calculations, we use initial states that consist of a compact triangle of 
alpha clusters and a fourth alpha cluster, located either in the plane of the triangle (square-like) or out of the plane of the triangle (tetrahedral).   

\begin{table}[t]
\centering
\begin{tabular}{c | c c | c}
& LO & rescaled & Exp \\ \hline\hline
$r(0^+_1)$ [fm] & 2.3(1) & --- & 2.710(15) \cite{Kim:1978st} \\
$r(0^+_2)$ [fm] & 2.3(1) & --- & --- \\
$r(2^+_1)$ [fm] & 2.3(1) & --- & --- \\
$Q(2^+_1)$ [$e$ fm$^2$] & 10(2) & 15(3) & --- \\
$B(E2, 2^+_1 \to 0^+_2)$ [$e^2$fm$^4$] & 22(4) & 46(8) & 65(7) \cite{Ajzenberg:1990a}\\
$B(E2, 2^+_1 \to 0^+_1)$ [$e^2$fm$^4$] & 3.0(7) & 6.2(1.6) & 7.4(2) \cite{Moreh:1985zz}\\
$M(E0, 0^+_2 \to 0^+_1)$ [$e$ fm$^2$] & 2.1(7) & 3.0(1.4) & 3.6(2) \cite{Miska:1975}
\end{tabular}
\caption{NLEFT results for the charge radius $r$, the quadrupole moment $Q$, and the electromagnetic transition amplitudes for $E2$ and 
$E0$ transitions, as defined in Ref.~\cite{Mottelson_book}.
We compare with empirical (Exp) values where these are known. For the quadrupole moment and the transition amplitudes, we also show
``rescaled'' LO results, which correct for the deviation from the empirical value of the charge radius at LO (see main text). 
The errors are one-standard-deviation estimates which include statistical Monte Carlo errors and 
uncertainties due to the extrapolation $N_t^{} \to \infty$. 
\label{oxygen2}}
\end{table}

In Table~\ref{oxygen2}, we note that the LO charge radius $r_{\text{LO}}^{}$ of the ground state of $^{16}$O is smaller than the 
empirical value $r_{\text{exp}}^{}$. This leads to a systematic deviation, which arises from the overall size of the second moment of the charge distribution. 
To compensate for this overall scaling mismatch, we have also calculated ``rescaled'' quantities multiplied by powers of the ratio 
$r_{\text{exp}}^{}/r_{\text{LO}}^{}$, according to the length dimension of each observable.
With such a scaling factor included, we find that the NLEFT predictions for the $E2$ and $E0$ transitions are in good agreement with 
available experimental values. 


\section{Conclusions and outlook}

We have presented an overview of the central NLEFT results for the low-lying even-parity spectra of $^{12}$C and $^{16}$O. This includes the
Hoyle state of $^{12}$C which plays a central role in the stellar nucleosynthesis of life-essential elements. We have also shown that the electromagnetic
properties and transition rates of $^{12}$C and $^{16}$O are in agreement with available experimental data. While the long-term objectives of NLEFT 
are to decrease the lattice spacing and include higher orders in the EFT expansion, we also find that the missing physics up to $^{28}$Si
can be approximated by an ``effective'' 4N interaction. These results represent an important step 
towards more comprehensive NLEFT calculations of medium-mass nuclei in the near future.

\acknowledgments

We are grateful for the help in automated data collection by Thomas Luu.
Partial financial support from the Deutsche Forschungsgemeinschaft (Sino-German CRC 110), the Helmholtz Association (Contract No.\ VH-VI-417), 
BMBF (Grant No.\ 05P12PDFTE), and the U.S. Department of Energy (DE-FG02-03ER41260) is acknowledged. This work was further supported
by the EU HadronPhysics3 project, and funds provided by the ERC Project No.\ 259218 NUCLEAREFT. The computational resources 
were provided by the J\"{u}lich Supercomputing Centre at the Forschungszentrum J\"{u}lich and by RWTH Aachen.



\begin{thebibliography}{99}

\bibitem{Hagen:2012fb} 
  G.~Hagen, M.~Hjorth-Jensen, G.~R.~Jansen, R.~Machleidt, and T.~Papenbrock,
  Phys.\ Rev.\ Lett.\  {\bf 109}, 032502 (2012).
  
\bibitem{Jurgenson:2013yya} 
  E.~D.~Jurgenson, P.~Maris, R.~J.~Furnstahl, P.~Navratil, W.~E.~Ormand and J.~P.~Vary,
  Phys.\ Rev.\ C {\bf 87}, 054312 (2013). 

\bibitem{Roth:2011ar} 
  R.~Roth, J.~Langhammer, A.~Calci, S.~Binder, and P.~Navratil,
  Phys.\ Rev.\ Lett.\ {\bf 107}, 072501 (2011).
  

\bibitem{Hergert:2012nb} 
  H.~Hergert, S.~K.~Bogner, S.~Binder, A.~Calci, J.~Langhammer, R.~Roth and A.~Schwenk,
  Phys.\ Rev.\ C {\bf 87}, 034307 (2013).

\bibitem{Lovato:2013cua} 
  A.~Lovato, S.~Gandolfi, R.~Butler, J.~Carlson, E.~Lusk, S.~C.~Pieper and R.~Schiavilla,
  Phys.\ Rev.\ Lett.\  {\bf 111}, 092501 (2013).

\bibitem{Duguet}
  V.~Som\`a, C.~Barbieri, and T.~Duguet,
  Phys.\ Rev.\ C {\bf 87}, 011303(R) (2013).

\bibitem{Epelbaum:2008ga} 
  E.~Epelbaum, H.-W.~Hammer, and Ulf-G.~Mei{\ss}ner,
  Rev.\ Mod.\ Phys.\  {\bf 81}, 1773 (2009).

\bibitem{Dean_QMC}
  D.~Lee,
  Prog.\ Part.\ Nucl.\ Phys.\ {\bf 63}, 117 (2009).

\bibitem{Borasoy:2006qn} 
  B.~Borasoy, E.~Epelbaum, H.~Krebs, D.~Lee, and Ulf-G.~Mei{\ss}ner,
  Eur.\ Phys.\ J.\ A {\bf 31}, 105 (2007).

\bibitem{Epelbaum:2009pd} 
  E.~Epelbaum, H.~Krebs, D.~Lee, and Ulf-G.~Mei{\ss}ner,
  Phys.\ Rev.\ Lett.\  {\bf 104}, 142501 (2010);
  {\it ibid.},
  Eur.\ Phys.\ J.\ A {\bf 45}, 335 (2010).


\bibitem{Epelbaum:2011md} 
  E.~Epelbaum, H.~Krebs, D.~Lee, and Ulf.-G.~Mei{\ss}ner,
  Phys.\ Rev.\ Lett.\  {\bf 106}, 192501 (2011).

\bibitem{Epelbaum:2012qn} 
  E.~Epelbaum, H.~Krebs, T.~A.~L\"ahde, D.~Lee, and Ulf-G.~Mei{\ss}ner,
  Phys.\ Rev.\ Lett.\ {\bf 109}, 252501 (2012);
  {\it ibid.},
  Phys.\ Rev.\ Lett.\ {\bf 110}, 112502 (2013);
  {\it ibid.},
  Eur.\ Phys.\ J.\ A {\bf 49}, 82 (2013).
  
\bibitem{A28_letter}
  T.~A.~L\"ahde, E.~Epelbaum, H.~Krebs, D.~Lee, Ulf-G.~Mei{\ss}ner, and G.~Rupak,
  arXiv:1311.0477 [nucl-th].

\bibitem{16O_spectrum}
  E.~Epelbaum, H.~Krebs, T.~A.~L\"ahde, D.~Lee, Ulf-G.~Mei{\ss}ner, and G.~Rupak,
  {\it to appear in Physical Review Letters},
arXiv:1312.7703 [nucl-th]

\bibitem{Oberhummer}
  H.~Schlattl, A.~Heger, H.~Oberhummer, T.~Rauscher, and A.~Cs\'ot\'o,
  Astrophys.\ Space Sci.\ {\bf 291}, 27 (2004).

\bibitem{WIMP}
 S.~R.~Beane, S.~D.~Cohen, W.~Detmold, H.-W.~Lin, and M.~J.~Savage,
 arXiv:1306.6939 [hep-ph].

\bibitem{Schaller:1982}
  L.~A.~Schaller {\it et al.},
  Nucl.\ Phys.\ {\bf A379}, 523 (1982).

\bibitem{Vermeer:1983}
  W.~J.~Vermeer {\it et al.},
  Phys.\ Lett.\ B {\bf 122}, 23 (1983).
  
\bibitem{Ajzenberg:1990a}
 F.~Ajzenberg-Selove,
 Nucl.\ Phys.\ {\bf B506}, 1 (1990).

\bibitem{Chernykh:2010zu}
  M.~Chernykh {\it et al.},
  Phys.\ Rev.\ Lett.\  {\bf 105}, 022501 (2010).



\bibitem{Wheeler:1937zza} 
  J.~A.~Wheeler,
  Phys.\ Rev.\ {\bf 52}, 1083 (1937).

\bibitem{Dennison:1954zz}
  D.~M.~Dennison,
  Phys.\ Rev.\ {\bf 96}, 378 (1954).
  
\bibitem{Iachello}
  H.~Feshbach and F.~Iachello,
  Phys.\ Lett.\ B {\bf 45}, 7 (1973).
  
\bibitem{Robson:1979zz}
  D.~Robson,
  Phys.\ Rev.\ Lett.\ {\bf 42}, 876 (1979).

\bibitem{Bauhoff:1984zza}
  W.~Bauhoff, H.~Schultheis, and R.~Schultheis,
  Phys.\ Rev.\ C {\bf 29}, 1046 (1984).

\bibitem{Filikhin:1999}
  I.~N.~Filikhin and S.~L.~Yakovlev,
  Phys.\ Atom.\ Nucl.\ {\bf 64}, 409 (2000).

\bibitem{Tohsaki:2001an}
  A.~Tohsaki, H.~Horiuchi, P.~Schuck, and G.~R\"opke,
  Phys.\ Rev.\ Lett.\ {\bf 87}, 192501 (2001).

\bibitem{Bijker}
  R.~Bijker, 
  AIP Conf.\ Proc.\ {\bf 1323}, 28 (2010);
  {\it ibid.}, J.\ Phys.:\ Conf.\ Ser.\ {\bf 380}, 012003 (2012).
  
\bibitem{Freer:2005ia}
  M.~Freer (CHARISSA Collaboration),
  J.\ Phys.\ {\bf G31}, S1795 (2005).




\bibitem{Mottelson_book}
  A.~Bohr and B.~R.~Mottelson,
  {\it Nuclear Structure. Single-Particle Motion}
  (W.~A.~Benjamin, New York, 1969), Vol.~I.

\bibitem{Kim:1978st}
  J.~C.~Kim {\it et al.},
  Nucl.\ Phys.\ {\bf A297}, 301 (1978).

  
\bibitem{Moreh:1985zz}
  R.~Moreh, W.~C.~Sellyey, D.~Sutton, and R.~Vodhanel,
  Phys.\ Rev.\ C {\bf 31}, 2314 (1985).

\bibitem{Miska:1975}
  H.~Miska {\it et al.},
  Phys.\ Lett.\ B {\bf 58}, 155 (1975).

\end{thebibliography}
\end{document}